\documentclass[dvips]{acta}
\usepackage{supertabular,lscape,epsfig}
\usepackage{amssymb}
\usepackage{amsmath}
\usepackage{float}
\mathchardef\mhyphen="2D
\usepackage{placeins}
\DeclareSymbolFont{ppa}{OT1}{ppl}{m}{it}
\DeclareMathSymbol{\vv}{\mathalpha}{ppa}{'166}

\SetPages{1}{12}

\SetVol{72}{2022}

\begin{document}
\newcommand\pvalue{\mathop{p\mhyphen {\rm value}}}
\newcommand{\TabApp}[2]{\begin{center}\parbox[t]{#1}{\centerline{
  {\bf Appendix}}
  \vskip2mm
  \centerline{\small {\spaceskip 2pt plus 1pt minus 1pt T a b l e}
  \refstepcounter{table}\thetable}
  \vskip2mm
  \centerline{\footnotesize #2}}
  \vskip3mm
\end{center}}

\newcommand{\TabCapp}[2]{\begin{center}\parbox[t]{#1}{\centerline{
  \small {\spaceskip 2pt plus 1pt minus 1pt T a b l e}
  \refstepcounter{table}\thetable}
  \vskip2mm
  \centerline{\footnotesize #2}}
  \vskip3mm
\end{center}}

\newcommand{\TTabCap}[3]{\begin{center}\parbox[t]{#1}{\centerline{
  \small {\spaceskip 2pt plus 1pt minus 1pt T a b l e}
  \refstepcounter{table}\thetable}
  \vskip2mm
  \centerline{\footnotesize #2}
  \centerline{\footnotesize #3}}
  \vskip1mm
\end{center}}

\newcommand{\MakeTableApp}[4]{\begin{table}[p]\TabApp{#2}{#3}
  \begin{center} \TableFont \begin{tabular}{#1} #4
  \end{tabular}\end{center}\end{table}}

\newcommand{\MakeTableSepp}[4]{\begin{table}[p]\TabCapp{#2}{#3}
  \begin{center} \TableFont \begin{tabular}{#1} #4
  \end{tabular}\end{center}\end{table}}

\newcommand{\MakeTableee}[4]{\begin{table}[htb]\TabCapp{#2}{#3}
  \begin{center} \TableFont \begin{tabular}{#1} #4
  \end{tabular}\end{center}\end{table}}

\newcommand{\MakeTablee}[5]{\begin{table}[htb]\TTabCap{#2}{#3}{#4}
  \begin{center} \TableFont \begin{tabular}{#1} #5
  \end{tabular}\end{center}\end{table}}

\newcommand{\MakeTableH}[4]{\begin{table}[H]\TabCap{#2}{#3}
  \begin{center} \TableFont \begin{tabular}{#1} #4
  \end{tabular}\end{center}\end{table}}

\newcommand{\MakeTableHH}[4]{\begin{table}[H]\TabCapp{#2}{#3}
  \begin{center} \TableFont \begin{tabular}{#1} #4
  \end{tabular}\end{center}\end{table}}

\newfont{\bb}{ptmbi8t at 12pt}
\newfont{\bbb}{cmbxti10}
\newfont{\bbbb}{cmbxti10 at 9pt}
\newcommand{\uprule}{\rule{0pt}{2.5ex}}
\newcommand{\douprule}{\rule[-2ex]{0pt}{4.5ex}}
\newcommand{\dorule}{\rule[-2ex]{0pt}{2ex}}
\def\thefootnote{\fnsymbol{footnote}}
\begin{Titlepage}
\Title{The OGLE Collection of Variable Stars.\\
Over 2600 $\delta$~Scuti Stars in the Small Magellanic Cloud\footnote{Based on observations obtained with the 1.3-m Warsaw telescope at the Las Campanas Observatory of the Carnegie Institution for Science.}}
\Author{I.~~S~o~s~z~y~{\'n}~s~k~i$^1$,~~
A.~~U~d~a~l~s~k~i$^1$,~~
J.~~S~k~o~w~r~o~n$^1$,~~
P.~~P~i~e~t~r~u~k~o~w~i~c~z$^1$,\\
M.\,K.~~S~z~y~m~a~{\'n}~s~k~i$^1$,~~
R.~~P~o~l~e~s~k~i$^1$,~~
D.\,M.~~S~k~o~w~r~o~n$^1$,~~
S.~~K~o~z~{\l}~o~w~s~k~i$^1$,\\
P.~~M~r~{\'o}~z$^1$,~~
P.~~I~w~a~n~e~k$^1$,~~
M.~~W~r~o~n~a$^1$,~~
K.~~U~l~a~c~z~y~k$^{2,1}$,~~
K.~~R~y~b~i~c~k~i$^{3.1}$,\\
and~~M.~~G~r~o~m~a~d~z~k~i$^1$
}
{$^1$Astronomical Observatory, University of Warsaw, Al.~Ujazdowskie~4, 00-478~Warszawa, Poland\\
$^2$Department of Physics, University of Warwick, Gibbet Hill Road, Coventry, CV4~7AL,~UK\\
$^3$Department of Particle Physics and Astrophysics, Weizmann Institute of Science, Rehovot 76100, Israel}
\Received{\today}
\end{Titlepage}

\Abstract{We present the first-ever collection of $\delta$~Scuti stars found over the entire area of the Small Magellanic Cloud (SMC). The sample consists of 2810 variables of which over 2600 objects belong to the SMC while the remaining stars are most likely members of the Milky Way's halo. The sample has been divided into 2733~singlemode and 77~multimode pulsators. We provide observational parameters (pulsation periods, mean magnitudes, amplitudes, Fourier coefficients) of all $\delta$~Sct stars and the long-term {\it I}- and {\it V}-band time-series photometric measurements collected during the fourth phase of the Optical Gravitational Lensing Experiment (OGLE-IV).}{Stars: variables: delta Scuti -- Stars: oscillations -- Magellanic Clouds -- Catalogs}

\vspace{0.5cm}
\Section{Introduction}

$\delta$~Scuti variables are intermediate-mass stars pulsating in low-order radial or non-radial modes driven by the $\kappa$ mechanism operating in the He~II ionization zone (Chevalier 1971). This class includes young and intermediate-age stars on the pre-main sequence, main sequence, and post-main sequence, as well as stars belonging to the old population that are probably merged binaries (Breger 2000). The latter group is referred to as SX~Phoenicis stars. The pulsation periods of $\delta$~Sct variables are shorter than 0.3~d, while the amplitudes reach 1~mag in the {\it V} band, although most stars of this type exhibit much smaller amplitudes, down to the sub-millimagnitude level (Balona and Dziembowski 2011).

$\delta$~Sct stars, like other classical pulsators, follow period--luminosity (PL) and period--luminosity--color (PLC) relations. The Large and Small Magellanic Cloud (LMC and SMC) play a crucial role in studying the PL relations obeyed by various types of pulsating stars (\eg Leavitt and Pickering 1912, Glass and Lloyd Evans 1981, Madore 1982, Udalski \etal 1999, Muraveva \etal 2015). The PL relations for $\delta$~Sct stars were also investigated based on variables detected in the Magellanic Clouds (\eg Poleski \etal 2010, McNamara 2011, Mart{\'\i}nez-V{\'a}zquez 2022), but these studies were limited by a small number of known $\delta$~Sct stars in the Clouds, especially in the SMC.

The first $\delta$~Sct variables in the SMC were discovered as a by-product of a search for RR~Lyr stars in the OGLE-II photometric database. Soszy{\'n}ski \etal (2002) published a list of 19 short-period variables (with periods in the range 0.16--0.57~d) suggesting that most of them are $\delta$~Sct stars. In the following years, about half of these stars were re-classified as RR~Lyr stars or Cepheids (see Section~4 for the definition of the boundary between classical Cepheids and $\delta$~Sct stars), but eight of them turned out to be {\it bona fide} $\delta$~Sct variables and are included in the present work. One more $\delta$~Sct star from the OGLE project was reported by Pietrukowicz (2018). The second, and so far the last, sample of $\delta$~Sct stars in the SMC was published by Mart{\'\i}nez-V{\'a}zquez \etal (2021). They used deep photometry of the SMC globular cluster NGC 419 obtained by the Gemini South telescope to detect 54 $\delta$~Sct stars of which six objects are probable cluster members and 48 are field variables. The Gaia DR3 catalog of main-sequence oscillators (Gaia Collaboration \etal 2022) contains over 50 $\delta$~Sct stars in the region of the sky occupied by the SMC, but all these pulsators belong to the Milky Way halo and are located in the foreground of the SMC.

Thus, only 63 $\delta$~Sct stars in the SMC have been known until now. With this paper, we release the first catalog of $\delta$~Sct variables found over the entire area of the SMC. Our sample consists of 2810 pulsators of which over 2600 objects are likely members of the SMC. This is part of the OGLE Collection of Variable Stars (OCVS), currently containing over a million manually classified variable stars in the Milky Way and Magellanic Clouds. In the next paper in this series, we will present the OGLE collection of over 14,000 $\delta$~Sct pulsators in the LMC.

This work is structured as follows. In Section~2, we present a summary of the OGLE survey of the Magellanic Clouds. Section~3 is devoted to a brief review of the variable star selection and classification procedures. In Section~4, we discuss the criteria that can be used to distinguish between $\delta$~Sct stars and classical Cepheids. The OGLE collection of $\delta$~Sct variables in the SMC is described in Section~5. The completeness of our catalog is discussed in Section~6. We close the paper with a summary in Section~7.

\Section{OGLE Observations of the Magellanic System}

The OGLE survey uses a dedicated 1.3-m Warsaw Telescope equipped with a large format mosaic CCD camera with a field of view of 1.4~square degrees and a pixel scale of 0.26~arcsec. The telescope is located at Las Campanas Observatory, Chile, operated by the Carnegie Institution for Science. Our collection of $\delta$~Sct stars is based on photometric data from the fourth phase of the OGLE project (OGLE-IV, Udalski \etal 2015), which has been operating since 2010 until today. In this work, we use the observations obtained before March 2020, when the Warsaw Telescope was temporarily closed due to the COVID-19 pandemic.

The OGLE observations are carried out in the {\it I} and {\it V} photometric bands, closely resembling those from the Cousins-Johnson standard systems. From several dozen to over 1000 observations per star (typically 600-700) have been acquired in the {\it I}-band and from several to over 300 (typically 40-60) data points have been gathered in the {\it V}-band light curves. The uniform exposure time of 150~s resulted in the saturation limit of the OGLE photometry at the level of about $I\approx13$~mag and the sensitivity limit at about $I\approx21.5$~mag.

The total sky area monitored by the OGLE survey in the region of the Magellanic System is 765 square degrees. The OGLE footprint covers both galaxies together with the Magellanic Bridge and vast regions on their periphery. For the purposes of this work, we adopted the celestial meridian of 2\zdot\uph8 as an arbitrary on-sky boundary between the LMC and SMC regions. The same value was selected to separate the LMC and SMC RR~Lyr stars in the OCVS (Soszy{\'n}ski \etal 2016). In the SMC area of about 280 square degrees, OGLE regularly observes about 14 million stars belonging to both the SMC and the halo of the Milky Way.

\Section{Selection and Classification of $\delta$~Sct Stars}

We began our search for $\delta$~Sct variables in the SMC with a massive period analysis for all 14~million {\it I}-band light curves stored in the OGLE-IV databases. We used the {\sc Fnpeaks} code\footnote{http://helas.astro.uni.wroc.pl/deliverables.php?lang=en\&active=fnpeaks} to span the frequency domain from 0 to 100~cycles per day with a step of $5\cdot10^{-5}$~cycles per day. For each light curve, the period with the highest signal-to-noise ratio was considered to be dominant. Then, this dominant frequency and its harmonics were subtracted from the light curve and the period search was repeated on the residual data.

\begin{figure}[t]
\includegraphics[bb = 45 160 570 760, clip, width=12.8cm]{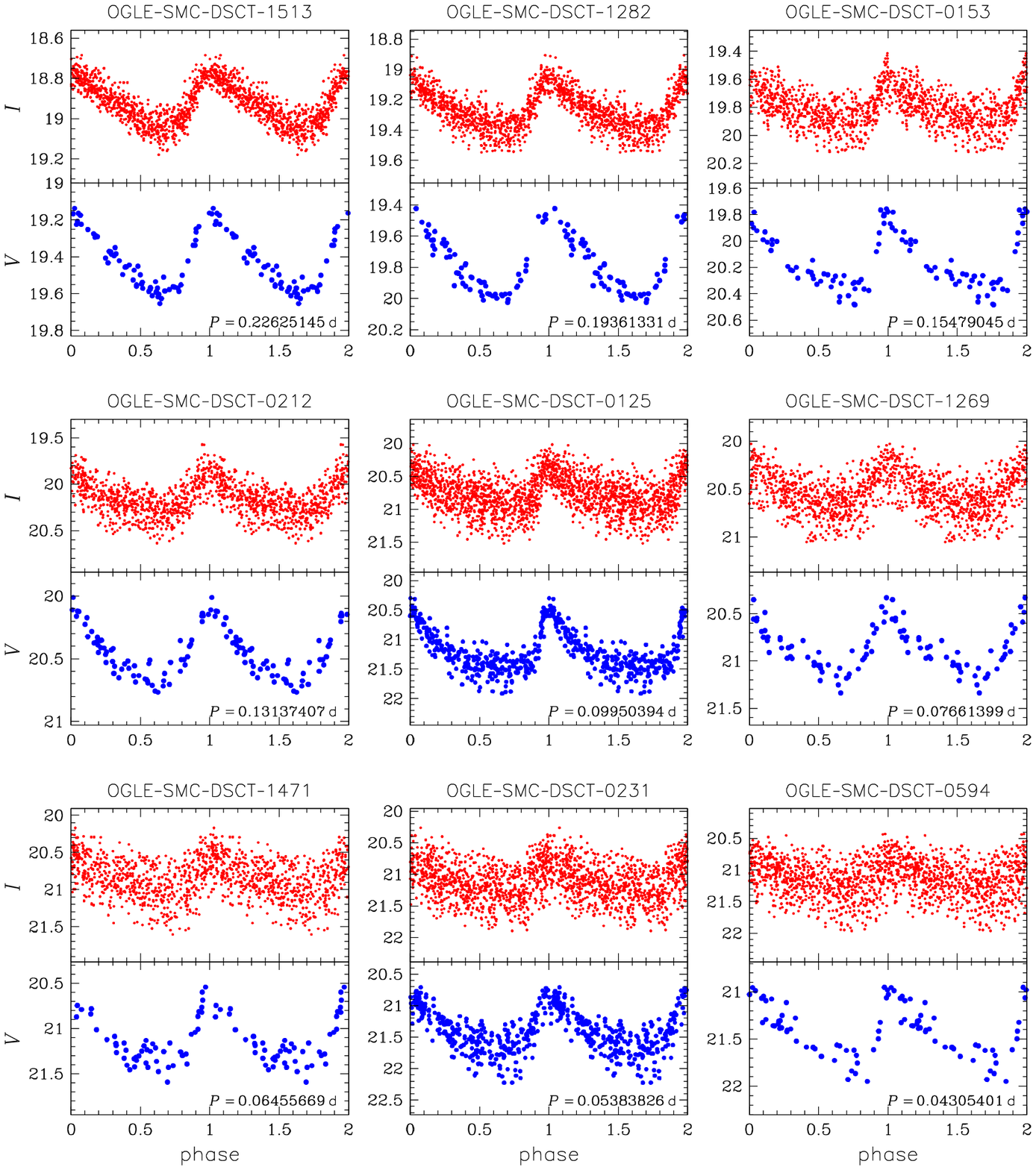}
\FigCap{OGLE-IV {\it I}-band (red) and {\it V}-band (blue) light curves of nine example $\delta$~Sct stars in the SMC.}
\end{figure}

In the next step of the procedure, we used both {\it I}- and {\it V}-band OGLE-IV photometric data to select and classify $\delta$~Sct stars in the SMC. The much smaller number of the OGLE observations in the {\it V}-band was compensated by the lower noise of the {\it V}-band light curves compared to the {\it I}-band ones. Fig.~1 shows the {\it I}- and {\it V}-band light curves of nine example $\delta$~Sct variables arranged in descending pulsation periods. Note that the shortest-period (and therefore the faintest) stars have a significant scatter of the data points in the {\it I}-band, while the {\it V}-band light curves of these objects are much less dispersed. However, every $\delta$~Sct variable identified based on the {\it V}-band data was later confirmed with the {\it I}-band observations, although in some cases the {\it I}-band light curves are very noisy.

Our variability classification was primarily based on the shapes of the {\it I}- and {\it V}-band light curves. For single-periodic variables, we filtered out stars with sinusoidal light curves and obvious eclipsing binaries. For multi-periodic objects, we took into account the characteristic period ratios of multimode pulsators.

\begin{figure}[t]
\includegraphics[width=12.5cm]{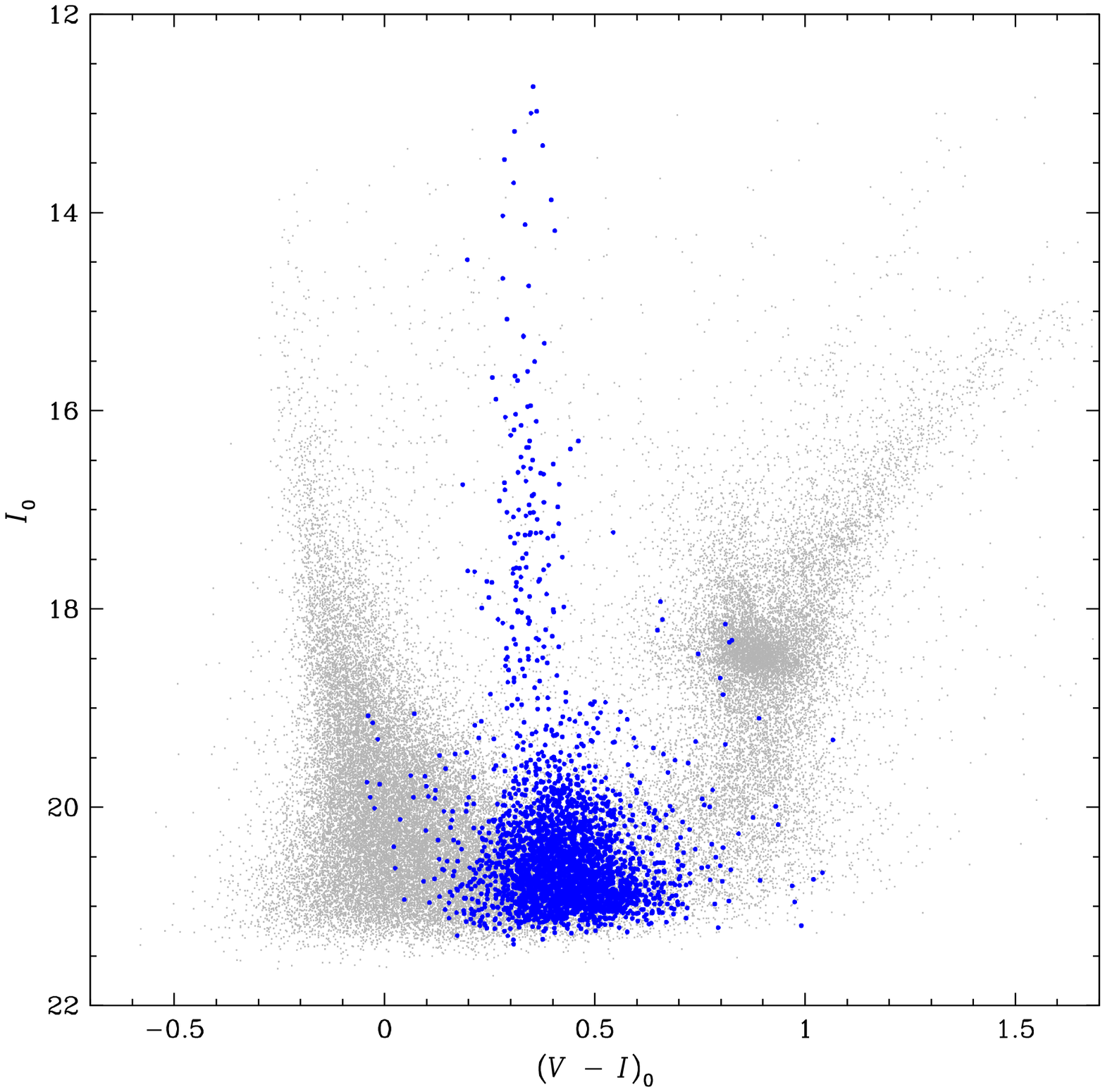}
\vspace*{2pt}
\FigCap{$(V-I)_0$ vs. $I_0$ color--magnitude diagram for $\delta$~Sct stars
in the SMC (blue points). The background gray points show stars from the subfield SMC719.11.}
\end{figure}

In the final stage of our classification procedure, we examined the positions of the candidate pulsation stars in the color--magnitude diagram depicted in Fig.~2, where the {\it I}-band magnitudes and $V-I$ color index were corrected for the interstellar extinction using the most accurate reddening maps of the Magellanic Clouds recently published by Skowron \etal (2021). Most of the stars with the dereddened color index $(V-I)_0<0.2$~mag or $(V-I)_0>0.7$~mag, \ie outside the $\delta$~Sct instability strip, were removed from the list. However, as can be seen in Fig.~2, we decided to keep some too-blue or too-red stars on the list, because their light curve morphology suggested that these objects could be genuine $\delta$~Sct variables blended with nearby unresolved stars. Nonetheless, the objects with atypical colors should be treated with caution as uncertain candidates for $\delta$~Sct stars.

Finally, our collection of $\delta$~Sct stars detected toward the SMC consists of 2810 objects. In addition, the OCVS has been enriched with nine new classical Cepheids and 123 RR Lyr stars identified as a by-product of the present search for $\delta$~Sct stars or a result of cross-matching the OGLE database with the Gaia DR3 catalog of Cepheids (Ripepi \etal 2022) and RR~Lyr stars (Clementini \etal 2022). A more detailed comparison of the Gaia DR3 and OGLE catalogs in the area of the Magellanic System will be discussed in the forthcoming paper presenting the OGLE collection of $\delta$~Sct stars in the LMC (Soszy{\'n}ski \etal in prep.).

\begin{figure}[p]
\includegraphics[bb = 37 50 540 745, clip, width=12.7cm]{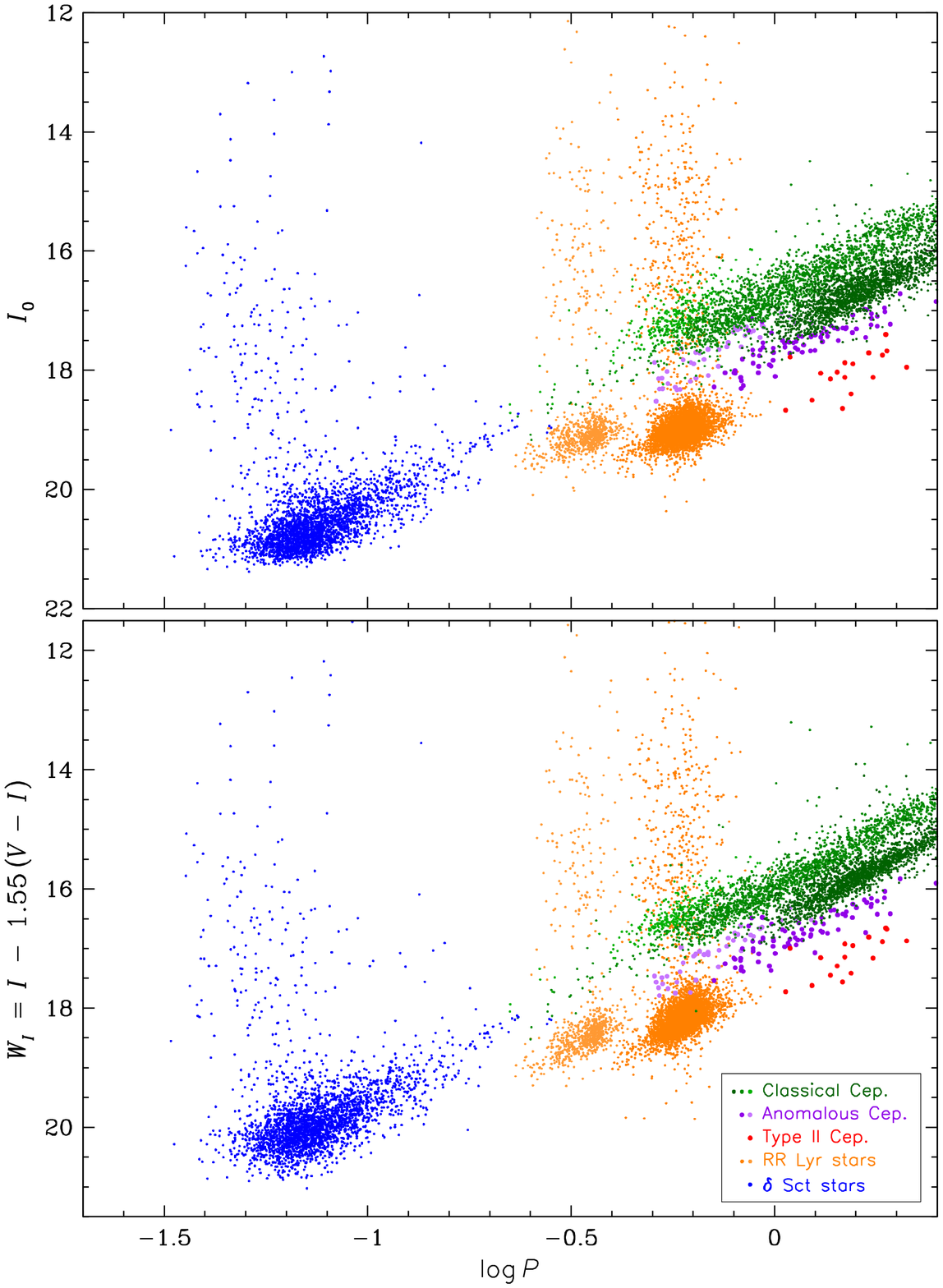}
\FigCap{Period--luminosity ({\it upper panel}) and period--Wesenheit index
({\it lower panel}) diagrams for classical pulsators in the SMC. Green
points show classical Cepheids, violet points -- anomalous Cepheids, red
points -- type II Cepheids, orange points -- RR Lyr stars, and blue points
-- $\delta$~Sct stars. Lighter colors indicate stars oscillating in higher
modes. Wesenheit index is a reddening-free function defined as
$W_I=I-1.55(V-I)$ (Madore 1982).}
\end{figure}

\Section{Boundary Between $\delta$~Sct Stars and Classical Cepheids}

Population I $\delta$~Sct stars and classical Cepheids occupy common sequences in the PL (Fig.~3), color--magnitude, period--radius (\eg Fernie 1992), Petersen (\eg Po\-leski \etal 2010) and other diagrams. In fact, it is a matter of convention what value of the pulsation period is adopted to separate both types of variable stars. Various authors proposed different maximum periods of $\delta$~Sct stars. For example, in the General Catalogue of Variable Stars (Samus' \etal 2017) it is 0.2~d, Breger (2000) defined $\delta$~Sct variables as pulsators with periods shorter than 0.25~d, Rodr{\'\i}guez \etal (2000) adopted 0.3~d, Handler (2009) -- 0.33~d (8~hr), while Catelan and Smith (2015) -- 0.42~d. 

One might ask if it is possible to establish a ``natural'' boundary period separating $\delta$~Sct stars from classical Cepheids. Both types of variables may pulsate, among others, in the fundamental mode (F), first-overtone (1O), second-overtone (2O), or in the mix of these radial modes. All but one of these combinations form continuity in the vicinity of the Cepheid--$\delta$~Sct border. The exception is a class of single-mode F pulsators that show a natural gap between $\delta$~Sct stars and classical Cepheids. OGLE-SMC-CEP-1899 ($P_\mathrm{F}=0.842$~d; Soszy{\'n}ski \etal 2010) is a classical Cepheids oscillating solely in the F mode with the shortest known period, while OGLE-LMC-DSCT-0617 ($P_\mathrm{F}=0.299$~d; Poleski \etal 2010) is the longest-period single-mode $\delta$~Sct star. Currently, no known Population~I single-mode F-mode star from the Cepheid instability strip pulsates with a period in the range of 0.3--0.8~d\footnote{Of course, the Population~II pulsators, like RR Lyr stars or anomalous Cepheids, may have periods in this range, but these stars can be isolated using other properties, \eg in the PL diagram, they lie below the classical Cepheid--$\delta$~Sct ridge.}. For this reason, all Population~I radially oscillating stars (both single- and multimode pulsators) with an F-mode period below 0.3~d are classified in our collection as $\delta$~Sct variables. Consequently, the 1O boundary period between $\delta$~Sct stars and Cepheids was fixed at $P_\mathrm{1O}=0.23$~d (corresponding to the period ratio $P_\mathrm{1O}/P_\mathrm{F}$ of about 0.77).

With the strict definition of different types of pulsating stars, we could extend the OCVS by nine short-period classical Cepheids in the SMC. One F/1O double-mode and eight 1O single-mode pulsators have 1O periods in the range 0.23--0.9~d. The OCVS (Soszy{\'n}ski \etal 2019) currently contains 4954 carefully selected classical Cepheids in the SMC.

\Section{The OGLE Collection of $\delta$ Sct Stars in the SMC}

Our collection of 2810 $\delta$~Sct variables in the SMC is available at the OGLE Internet Archive:
\begin{center}
{\it https://ogle.astrouw.edu.pl $\rightarrow$ OGLE Collection of Variable Stars}\\
{\it https://www.astrouw.edu.pl/ogle/ogle4/OCVS/smc/dsct/}\\
\end{center}

The sample was divided into 2733 singlemode and 77~multimode pulsators. The latter group contains objects with well-defined secondary or tertiary periods that may be associated with additional radial or non-radial pulsation modes. At least 160 stars in our sample are members of our Galaxy -- these are likely SX~Phe variables in the halo of the Milky Way located in front of the SMC. Fig.~4 shows on-sky maps of $\delta$~Sct stars from our collection. The left panel presents the spatial distribution of probable members of the SMC, while the right panel shows the positions of the putative Milky Way SX~Phe stars. The line separating these two populations was arbitrarily set at 1.5 mag above the mean PL relation for the SMC $\delta$~Sct variables (Fig.~3). More sophisticated methods would probably divide the Galactic and SMC populations more efficiently, however, one should remember that this separation cannot be done unequivocally because the Magellanic Clouds are submerged in the Milky Way halo -- old stellar populations in both galaxies smoothly transition into each other.

The full list of $\delta$~Sct stars with their equatorial coordinates (J2000.0), classification (singlemode, multimode), internal designations in the OGLE-IV, OGLE-III, and OGLE-II databases (if available), and the cross-matches with the International Variable Star Index (VSX; Watson \etal 2006) are provided in the {\sf ident.dat} file. In total, we identified only 11 common stars between our collection and the VSX catalog -- all of them are unquestionable members of the Milky Way.

The file {\sf dsct.dat} contains observation parameters of the stars: their intensity-averaged mean magnitudes in the {\it I}- and {\it V}-bands, up to three pulsation periods derived with the {\sc Tatry} code (Schwarzenberg-Czerny 1996), epochs of the maximum light, {\it I}-band peak-to-peak amplitudes, and Fourier coefficients. The periods are very precisely determined thanks to the long timespan of the OGLE-IV light curves, however, it cannot be ruled out that in the case of some the most noisy light curves our approach produced daily aliases of the true periods.

The OGLE-IV time-series photometry in the {\it I}- and {\it V}-bands are stored in the directory {\sf phot/}. Obvious outlying points (deviating by more than $\pm4$~sigma from the fitted model) were removed from the light curves, but it was verified beforehand whether these points are due to, for example, additional eclipses. We did not find any $\delta$~Sct stars with eclipses in the SMC. Finding charts can be found in the directory {\sf fcharts/}. These are $60''\times60''$ subframes of the {\it I}-band reference images, oriented with North at the top and East to the left.

\begin{landscape}
\begin{figure}[p]
\centerline{\includegraphics[width=20.5cm]{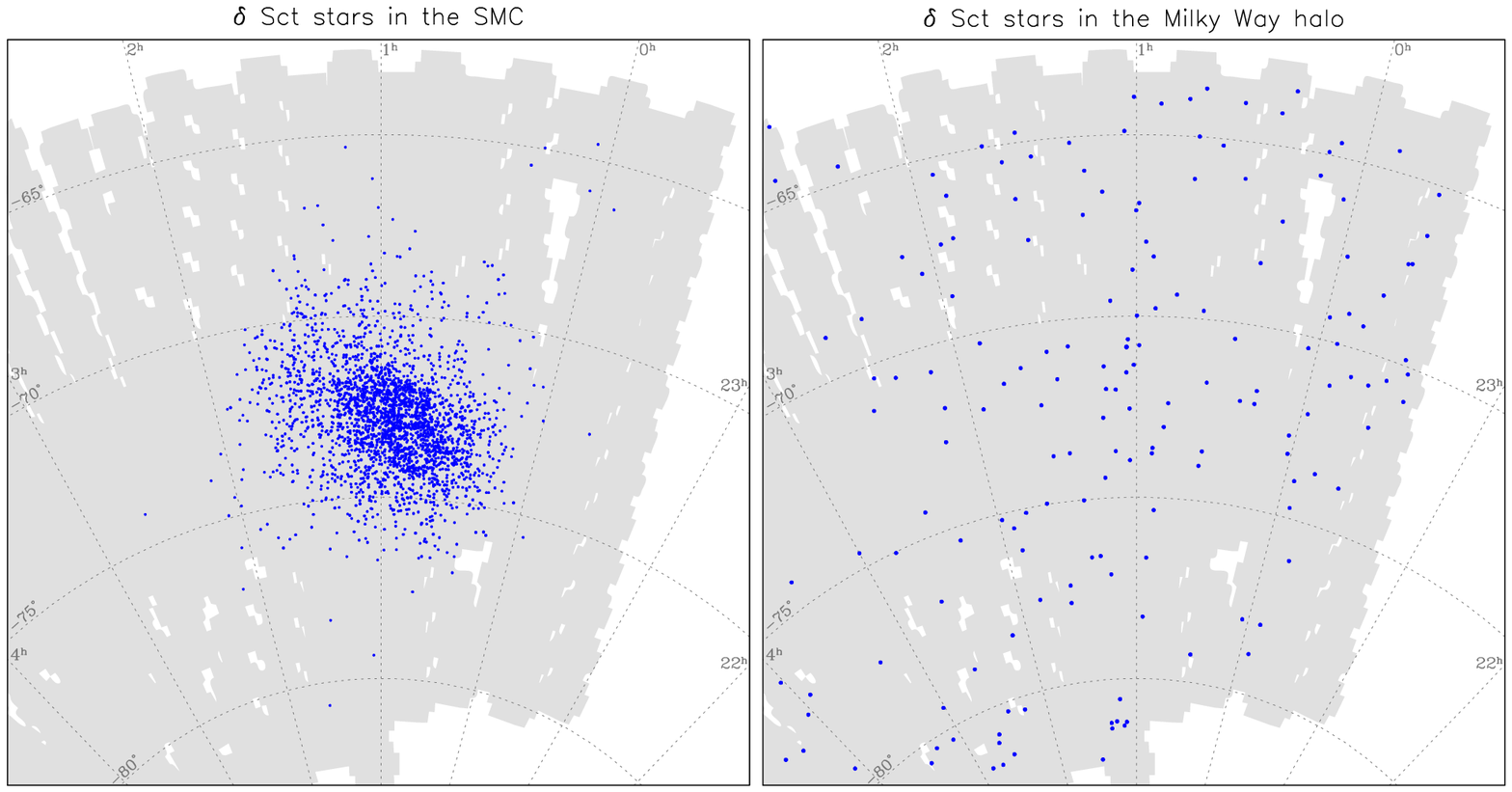}}
\vskip9pt
\FigCap{On-sky distributions of $\delta$~Sct stars toward the SMC. {\it Left
panel} shows positions of over 2600 probable members of the SMC. {\it Right
panel} shows $\delta$~Sct variables that likely belong to the Milky Way
halo. The gray area shows the OGLE footprint in the Magellanic System region.}
\end{figure}
\end{landscape}

\Section{Completeness of the Catalog}

The completeness of our collection is strongly limited by the brightness, amplitudes, and light curve shapes of $\delta$~Sct stars. Figs.~2 and~3 show that the number of variables in our sample drops sharply for mean magnitudes $I>21.2$~mag due to the sensitivity limit of the OGLE photometry.

We cross-matched our collection with the list of 54 $\delta$~Sct variables discovered with the Gemini South telescope in the field surrounding the SMC globular cluster NGC~419 (Mart{\'\i}nez-V{\'a}zquez \etal 2021). We found only one common object in both catalogs -- V46 = OGLE-SMC-DSCT-2067 -- which is the brightest $\delta$~Sct star detected by Mart{\'\i}nez-V{\'a}zquez \etal (2021). The remaining $\delta$~Sct variables identified with the 8.1-meter Gemini telescope are below the detection limit of the 1.3-meter Warsaw telescope used by the OGLE survey.

\begin{figure}[t]
\includegraphics[width=12.7cm]{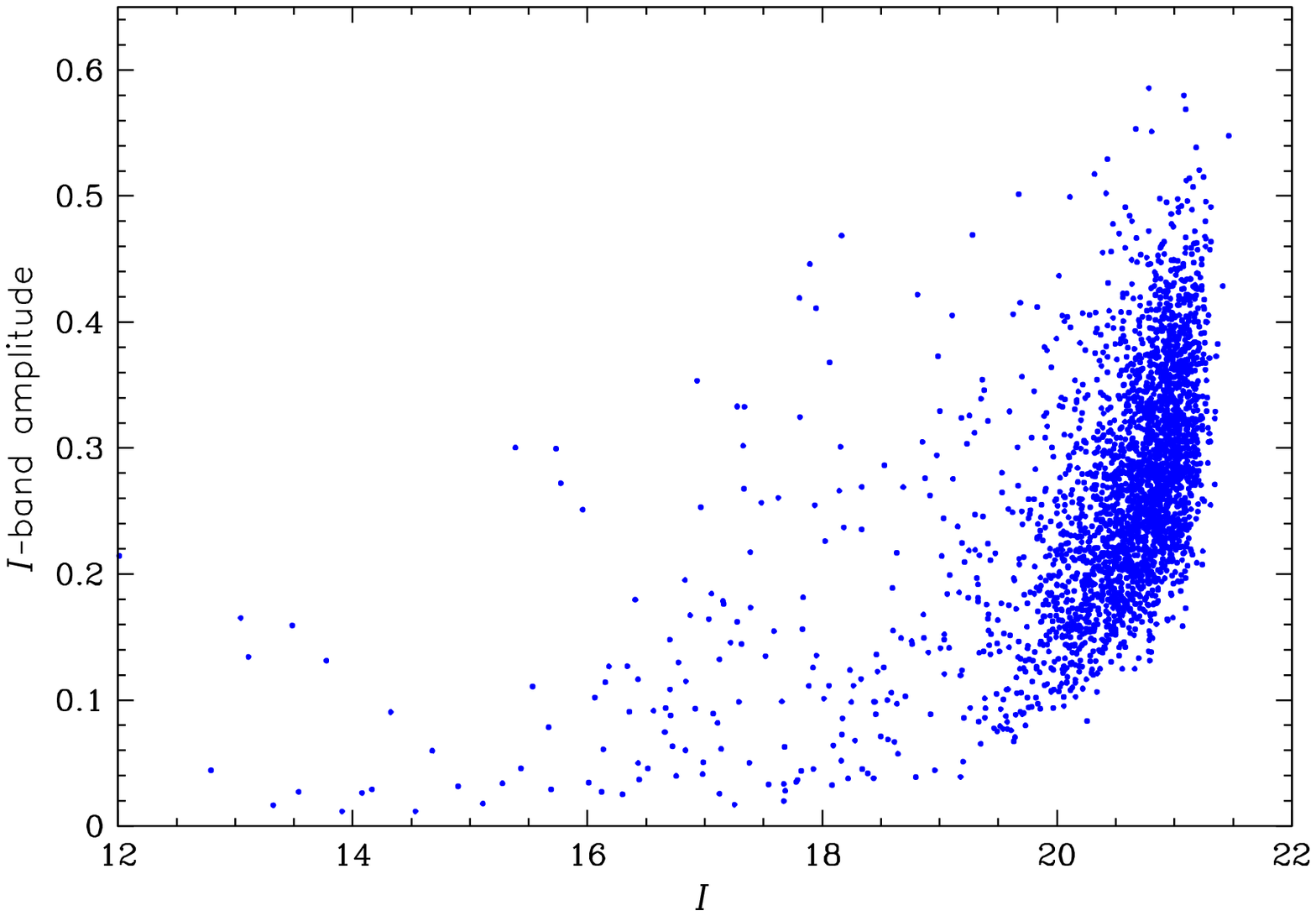}
\FigCap{Luminosity--amplitude diagram for $\delta$~Sct stars in the SMC.}
\end{figure}

The vast majority of $\delta$~Sct variables are small-amplitude pulsators (Breger 2000, Balona and Dziembowski 2011), so our collection must be the tip of the iceberg because we had the opportunity to discover only the high-amplitude $\delta$~Sct stars in the SMC. Fig.~5 shows the {\it I}-band peak-to-peak amplitudes plotted against the mean {\it I}-band magnitudes of our sample. It is obvious that the amplitude detection limits in the OGLE data are strongly correlated with the brightness of the pulsating candidates: for stars with the luminosities of about $I=19$~mag, the smallest detectable amplitudes are of the order of 0.04~mag, for $I=20$~mag stars, the amplitude detection limit rises to 0.1~mag, and for the faintest stars in our collection ($I=21$~mag), the light curve amplitude should be larger than about 0.2~mag to be recognized as a $\delta$~Sct variable. On the other hand, our sample of the Galactic $\delta$~Sct stars ($I<18.5$~mag) should be much more complete because we could detect variables with amplitudes as small as 0.01~mag in the {\it I}-band.

We used the $\delta$~Sct stars with double entries in the OGLE-IV database to estimate the completeness of our collection within the brightness and amplitude detection limits. These objects are located in the overlapping parts of the adjacent OGLE fields, so we had the opportunity to identify them twice during the selection and classification process. We found that 311 out of 2810 $\delta$~Sct stars included in the final version of our collection have their duplicates in the OGLE-IV database, so we had a chance to detect 622 counterparts. In fact, 449 objects from this list were independently classified as $\delta$~Sct variables which correspond to the catalog completeness slightly above 60\%. However, if we consider only stars brighter than $I=19$~mag, the formal completeness of our collection increases to about 85\%. Nonetheless, one should remember that the bulk of $\delta$~Sct stars in the SMC have luminosities and amplitudes below the detection limits of the OGLE project.

\Section{Summary}

We presented the first-ever catalog of $\delta$~Sct stars found in the entire area of the SMC. The number of known $\delta$~Sct variables in this galaxy has increased from about 60 to over 2600. Additionally, our collection contains at least 160 Galactic SX~Phe stars located in the foreground of the SMC. The most obvious application of our collection includes the examination of the PL and PLC relations obeyed by $\delta$~Sct stars in the metal-poor environment of the SMC compared to the more metal-rich variables in the LMC and Milky Way. The three-dimensional distribution of $\delta$~Sct variables can be analyzed to better understand the structure of the SMC. Moreover, the long-term photometric time series produced by the OGLE project can be used to study the stability of stellar pulsations over long time scales.

This work demonstrates the great versatility of OGLE photometric data. The OCVS contains variable stars with periods ranging from minutes to years, luminosities from 12~mag to over 22~mag in the {\it V}-band, and amplitudes from millimagnitudes to several magnitudes. We expect the next breakthrough in the field of variable stars in the Magellanic Clouds to come after the launch of the LSST project on the Rubin Observatory (Hambleton \etal 2022).

\Acknow{This work has been supported by the National Science Centre, Poland, grant no.~2022/45/B/ST9/00243. MG is supported by the EU Horizon 2020 research and innovation programme under grant agreement no.~101004719. This research has made use of the International Variable Star Index (VSX) database, operated at AAVSO, Cambridge, Massachusetts, USA.}


\begin{references}
\refitem{Balona, L.A., and Dziembowski, W.A.}{2011}{\MNRAS}{417}{591} 2011MNRAS.417..591B
\refitem{Breger M.}{2000}{~}{~}{in Breger M., Montgomery M., eds, ASP Conf. Ser. Vol. 210, Delta Scuti and Related Stars. Astron. Soc. Pac., San Francisco, p. 3}
\refitem{Catelan, M., and Smith, H.A.}{2015}{~}{~}{Pulsating Stars (New York: Wiley)} 
\refitem{Chevalier, C.}{1971}{\AA}{14}{24} 
\refitem{Clementini, G., Ripepi, V., Garofalo, A., \etal}{2022}{~}{~}{arXiv:2206.06278}
\refitem{Fernie, J.D.}{1992}{\AJ}{103}{1647}
\refitem{Gaia Collaboration, De Ridder, J., Ripepi, V., \etal}{2022}{~}{~}{arXiv e-prints, arXiv:2206.06075} 
\refitem{Glass, I.S., and Lloyd Evans, T.}{1981}{Nature}{291}{303}
\refitem{Hambleton, K.M., Bianco, F.B., Street, R., \etal.}{2022}{~}{~}{arXiv:2208.04499}
\refitem{Handler, G.}{2009}{~}{~}{in AIP Conf. Ser. 1170, Stellar Pulsation: Challenges for Theory and Observation, ed. J. A. Guzik \& P. A. Bradley (Melville, NY: AIP), 403}
\refitem{Leavitt, H.S., and Pickering, E.C.}{1912}{ Harvard Coll. Obs. Circ.}{173}{1}
\refitem{Madore, B.F.}{1982}{\ApJ}{253}{575}
\refitem{Mart{\'\i}nez-V{\'a}zquez, C.E., Salinas, R., and Vivas, A.K.}{2021}{\AJ}{161}{120}
\refitem{Mart{\'\i}nez-V{\'a}zquez, C.E., Salinas, R., Vivas, A.K., and Catelan, M.}{2022}{\ApJ}{940}{L25}
\refitem{McNamara, D.H.}{2011}{\AJ}{142}{110}
\refitem{Muraveva, T., Palmer, M., Clementini, G., \etal}{2015}{\ApJ}{807}{127}
\refitem{Pietrukowicz, P.}{2018}{~}{~}{in The RR Lyrae 2017 Conference. Revival of the Classical Pulsators: from Galactic Structure to Stellar Interior Diagnostics., Vol. 6., ed. R. Smolec, K. Kinemuchi, \& R.I. Anderson (Poland: Proc. Polish Astronomical Soc.), 258}
\refitem{Poleski, R., Soszy{\'n}ski, I., Udalski, A., \etal}{2010}{\Acta}{60}{1}
\refitem{Ripepi, V., Clementini, G., Molinaro, R., \etal}{2022}{~}{~}{arXiv:2206.06212}
\refitem{Rodr{\'\i}guez, E., L{\'o}pez-Gonz{\'a}lez, M.J., and L{\'o}pez de Coca, P.}{2000}{\AAS}{144}{469}
\refitem{Samus', N.N., Kazarovets, E.V., Durlevich, O.V., Kireeva, N.N., and Pastukhova, E.N.}{2017}{Astronomy Reports}{61}{80}
\refitem{Schwarzenberg-Czerny, A.}{1996}{\ApJ}{460}{L107}
\refitem{Skowron, D.M., Skowron, J., Udalski, A., \etal}{2021}{\ApJS}{252}{23}
\refitem{Soszy{\'n}ski, I., Udalski, A., Szyma{\'n}ski, M., \etal}{2002}{\Acta}{52}{369}
\refitem{Soszy{\'n}ski, I., Poleski, R., Udalski, A., \etal}{2010}{\Acta}{60}{17}
\refitem{Soszy{\'n}ski, I., Udalski, A., Szyma{\'n}ski, M.K., \etal}{2016}{\Acta}{66}{131}
\refitem{Soszy{\'n}ski, I., Udalski, A., Szyma{\'n}ski, M.K., \etal}{2019}{\Acta}{69}{87}
\refitem{Udalski, A., Szyma{\'n}ski, M., Kubiak, M., Pietrzy{\'n}ski, G., Soszy{\'n}ski, I., Wo{\'z}niak, P., and {\.Z}ebru{\'n}, K.}{1999}{\Acta}{49}{201}
\refitem{Udalski, A., Szyma{\'n}ski, M.K., and Szyma{\'n}ski, G.}{2015}{\Acta}{65}{1}
\refitem{Watson, C.L., Henden, A.A., and Price, A.}{2006}{Soc. Astron. Sci. Annu. Symp.}{25}{47}
\end{references}
\end{document}